\documentclass[sigconf,screen,nonacm]{acmart}

\AtBeginDocument{%
  }

\usepackage{algorithmic}
\usepackage{graphicx}
\usepackage{textcomp}
\usepackage{xcolor}
\usepackage{url}
\usepackage{pifont}

\usepackage{booktabs}
\usepackage{arydshln}
\setcopyright{acmlicensed}
\copyrightyear{2018}
\acmYear{2018}
\acmDOI{XXXXXXX.XXXXXXX}
\acmConference[Conference acronym 'XX]{Make sure to enter the correct
  conference title from your rights confirmation email}{June 03--05,
  2018}{Woodstock, NY}
\acmISBN{978-1-4503-XXXX-X/2018/06}

\newcommand*\rot{\rotatebox{90}}
\newcommand{\ok}[1]{\ding{51}$_{\text{#1}}$}

\begin{document}

\title{Benchmarking Practices in LLM-driven Offensive Security:
Testbeds, Metrics, and Experiment Design}

\author{Andreas Happe}
\email{andreas.happe@tuwien.ac.at}
\orcid{0009-0000-2484-0109}
\affiliation{%
  \institution{TU Wien}
  \city{Vienna}
  \country{Austria}
}

\author{Jürgen Cito}
\email{juergen.cito@tuwien.ac.at}
\orcid{0000-0001-8619-1271}
\affiliation{%
  \institution{TU Wien}
  \city{Vienna}
  \country{Austria}
}


\begin{abstract}
Large Language Models (LLMs) have emerged as a powerful approach for driving offensive penetration-testing tooling. Due to the opaque nature of LLMs, empirical methods are typically used to analyze their efficacy. The quality of this analysis is highly dependent on the chosen testbed, captured metrics and analysis methods employed.

This paper analyzes the methodology and benchmarking practices used for evaluating Large Language Model (LLM)-driven attacks, focusing on offensive uses of LLMs in cybersecurity. We review 19 research papers detailing 18 prototypes and their respective testbeds.

We detail our findings and provide actionable recommendations for future research, emphasizing the importance of extending existing testbeds, creating baselines, and including comprehensive metrics and qualitative analysis. We also note the distinction between security research and practice, suggesting that CTF-based challenges may not fully represent real-world penetration testing scenarios.
\end{abstract}



\maketitle

\section{Introduction}

The rapid evolution of Large Language Models (LLMs) has led to remarkable capabilities in various tasks, including offensive security tasks such as penetration testing, vulnerability discovery and exploitation~\cite{hassanin2024comprehensiveoverviewlargelanguage, YAO2024100211, jin2024llmsllmbasedagentssoftware, 10613562, xu2024largelanguagemodelscyber, motlagh2024largelanguagemodelscybersecurity, dube2024large, 10.1145/3659677.3659749, yigit2024reviewgenerativeaimethods, zhang2024llmsmeetcybersecuritysystematic}. 
The opaque nature of LLMs requires employing empirical methods for their evaluation. Thus, security researchers investigating the use of LLMs for offensive security depend on benchmarking and testbeds to assess the efficacy and effectiveness of their respective prototypes.

This paper provides an empirical investigation of testbeds used within offensive security research. We investigate their capabilities as well as metrics captured during prototype evaluation. We detail our findings and provide actionable recommendations for future research. Given the substantial costs of performing experiments using reasoning LLMs, we believe that this paper will offer valuable insights regarding experiment design for future publications.

We focus on testbeds for offensive use of LLMs, e.g., using LLMs for red-teaming~\cite{happe2023getting}. We do not analyze testbeds for testing the security of LLMs themselves such as CyberSecEval~\cite{bhatt2023purple,bhatt2024cyberseceval,wan2024cyberseceval}, i.e., red-teaming LLMs.

\subsection{Contributions}

\begin{table*}[ht]
\caption{CTF platforms mentioned within reviewed publications. For each platform, its full name and, if existing, a commonly used abbreviation is given. We differentiate between platforms offering VMs or cloud-based offerings. The former allow for local execution and archival of challenges; the latter are typically ``walled-gardens'' accessible within the cloud.}
\begin{center}
\begin{tabular}{lllll}
\toprule
\textbf{Abbrev.} & \textbf{Name} & \textbf{VM/Cloud} & \textbf{Description} & \textbf{Source}\\
 \midrule
 THM& TryHackMe & Cloud & Educational CTF platform & \url{https://tryhackme.com}\\\hdashline
 HTB& HackTheBox & Cloud & Educational CTF platform & \url{https://www.hackthebox.com/}\\\hdashline
 &picoCTF & Cloud & CMU CTF education platform & \url{https://picoctf.org/} \\\hdashline
 &lin.security & VM & linux privesc VM & \url{https://www.vulnhub.com/entry/linsecurity-1,244/}\\\hdashline
 &metasploitable2 & VM & metasploit education VM & \url{https://docs.rapid7.com/metasploit/metasploitable-2/} \\\hdashline
 OTW & OverTheWire & Cloud & Educational CTF challenges & \url{https://overthewire.org/wargames/} \\\hdashline
 & VulnHub & VM & vulnerable VM collection & \url{https://www.vulnhub.com/} \\\hdashline
 GOAD & A Game of AD & VM & Educational vulnerable AD & \url{https://github.com/Orange-Cyberdefense/GOAD} \\
 \bottomrule
\end{tabular}
\label{table:ctfs}
\end{center}
\end{table*}

Our motivation is to present current common practices with regard to testbeds, benchmarks, metrics and employed analysis techniques. Based upon the identified common practices, we identify sensible recommendation to which future research can depend to. To the best of our knowledge this publication is the first that provides an empirical investigation for the following relevant areas for researchers:

\begin{itemize}
    \item Composition and provenance of testbeds used for evaluating offensive LLMs.
    \item Experiment design including guidance on used metrics, sample sizes, and LLM selection.
    \item Methods used for analysis.
\end{itemize}

\section{Background}
\label{background}

The background section establishes a common understanding of used terms and subsequently provides background information about common penetration testing standards and CTF challenges referenced later within this work.

\subsection{Definitions: Testbeds, Baselines and Benchmarks}

According to Merriam-Webster, a testbed is ``\textit{any device, facility, or means for testing something in development}'' while a baseline is a ``\textit{a usually initial set of critical observations or data used for comparison or a control}.'' A benchmark is defined by ``\textit{something that serves as a standard by which others may be measured or judged}'' or by ``\textit{a standardized problem or test that serves as a basis for evaluation or comparison}''. The former can be achieved by using a baseline as benchmark, the latter can be achieved if a testbed consists of multiple atomic test-cases for which the test subjects success rate can be measured.

\subsection{Penetration Testing Standards}

Research indicates that penetration tests are not standardized for all domains, or that security professionals do not heed documented standards~\cite{happe2023understanding, van2019standardised}. Attack methodologies such as NIST 800-115~\cite{10.5555/2206199} or the Lockheed Martin Cyber Kill Chain detail different attack phases, not concrete attacks~\cite{10005490}. For example, the Cyber Kill Chain includes the following phases: Reconnaissance, Weaponization, Delivery, Exploitation, Installation, Command and Control, and Actions on Objective.

Real-life penetration tests are often structured around ``Top 10'' vulnerability lists for various areas~\cite{happe2023understanding}, e.g., the OWASP Top 10 for commonly used web vulnerabilities. The included Top 10 items are often broad and do not provide authoritative test cases. For example, the OWASP Top 10 contains the entry ``Injection'' that could be achieved through dozens of attack techniques and procedures. Another example of a ``fuzzy'' Top 10 item is ``Security Misconfiguration''.

Taxonomies such as MITRE ATT\&CK provide detailed information about attackers' techniques and tooling without providing overall attack strategies~\cite{10005490}.

Penetration Testing and its employed techniques is continuously changing. For example, the renowned OSCP certification\footnote{\url{https://www.offsec.com/courses/pen-200/}} changed its focus from exploit writing, e.g., creating buffer overflow exploits, to include more web vulnerabilities as well as Active Directory exploitation.

\subsection{CTF Challenges}

Our investigated testbeds commonly include tasks based on Capture-the-Flag (CTF) challenges in which the player typically has to exploit one or multiple vulnerabilities to gather a flag (secret string) as proof of compromise. CTFs typically include a diverse set of tasks, including cryptography, steganography, forensics, logic ``puzzles'', exploitation writing, privilege escalation, network attacks and web exploitation challenges. CTFs are often used for educational purposes, e.g., for training new security professionals. Empirical research has shown that they support knowledge transfer~\cite{happe2023understanding,kaplan2022capture,karagiannis2020analysis}, i.e., patterns and techniques learned during CTF exercises can be applied during real-world penetration-testing assignments.

CTFs can be classified into Jeopardy and Attacker/Defender challenges. In Jeopardy-style CTFs, participants face a series of separate challenges categorized into their respective topics. They are easier to score and analyze, but offer reduced realism. In Attacker/Defender CTFs, participants have to defend their infrastructure while attacking other teams' infrastructure. They typically employ simulated networks with vulnerable systems. These challenges offer more realistic scenarios but are complex to organize and require additional resources. Jeopardy-style challenges are often used for educational events that need to scale-out for many participants, while Attacker/Defender-style challenges are typically more advanced team-oriented events such as the NATO Locked Shields exercise\footnote{\url{https://ccdcoe.org/locked-shields/}}.

Table~\ref{table:ctfs} gives an overview of CTF platforms mentioned within our reviewed papers. While the mentioned cloud-based CTFs are free to use or provide free tiers, they do not make the building instructions of their challenges available publicly, and thus cannot be reproduced locally.

\section{Methodology}

We used Google Scholar to identify surveys containing the keywords ``offensive security LLM'' (\cite{hassanin2024comprehensiveoverviewlargelanguage, YAO2024100211, jin2024llmsllmbasedagentssoftware, 10613562, xu2024largelanguagemodelscyber, motlagh2024largelanguagemodelscybersecurity, dube2024large, 10.1145/3659677.3659749, yigit2024reviewgenerativeaimethods, zhang2024llmsmeetcybersecuritysystematic}). We analyzed surveyed publications and limited our selection to English publications released between 2023--2025 with a cut-off date of May 2025.

Publications had to include both an LLM-driven prototype for penetration testing as well as an empirical evaluation of their prototype using a documented testbed. We performed exponential non-discriminative snowball sampling (forward-referencing) by including papers linked from our initial paper seed, resulting in our final 19 papers detailing 18 prototypes and their respective testbeds detailed in Table~\ref{table:papers}.\footnote{The discrepancy between selected publications and testbeds results from two papers detailing the NYU CTF and respective offensive attack prototype. One paper details the testbed while the other paper details the offensive prototype.} Using forward-referencing also reduces the internal threat of selection bias.

We performed multi-stage thematic analysis~\cite{braun2006using,robson2002real}. Initially, each author read the gathered papers and identified themes. To increase trustworthiness~\cite{nowell2017thematic}, \textit{reflexive journaling} was employed and the identified themes discussed with two professional penetration-testers (\textit{peer debriefing}~\cite{janesick2007peer}). Subsequently, \textit{team consensus}~\cite{nowell2017thematic} was used to create the final themes. Subsequently, all papers were coded using the refined themes, resulting in the data used within this paper. 

\begin{table*}[ht]
\caption{Publications included in this survey. \textit{Initial Version} and \textit{Current Version} indicate the first and latest version of the publication available on arXiv. \textit{Versions} (\textit{V.})gives the number of paper versions available on arXiv. If a paper has already been published, the used publication outlet is indicated by \textit{Venue}, \textit{WS} indicates a workshop. Publications are listed in chronological order as given by the date of initial publication on arXiv.}
\begin{center}
\begin{tabular}{p{6.5cm}lllll}
\toprule
Publication & Authors & \shortstack[l]{Initial\\ Version} & V.& \shortstack[l]{Current\\ Version} & Venue \\
\midrule
Getting pwned by AI~\cite{happe2023getting} & Happe et al. & 2023-07-24 & 3 & 2023-08-17 & ESEC/FSE'23 \\\hdashline
PentestGPT~\cite{deng2024pentestgpt} & Deng et al. & 2023-08-13 & 2 & 2024-06-02 & Usenix Security'24 \\\hdashline
LLMs as Hackers~\cite{happe2024llms} & Happe et al. & 2023-10-17 & 5 & 2025-02-18 & \\\hdashline
Llm agents can autonomously hack websites~\cite{fang2024llmagentsautonomouslyhack} & Fang et al. & 2024-02-06 & 3 & 2024-06-16 & \\\hdashline
An empirical eval. of llms for solving offensive security challenges~\cite{shao2024empiricalevaluationllmssolving} & Shao et al. & 2024-02-19 & & \\\hdashline
AutoAttacker~\cite{xu2024autoattacker} & Xu et al. & 2024-03-02 & & & \\\hdashline
Llm agents can autonom. exploit one-day vulns.~\cite{fang2024llmagentsautonomouslyexploit} & Fang et al. & 2024-04-11 & 2 & 2024-04-17 & \\\hdashline
Teams of llm agents can exploit zero-day vulns.~\cite{fang2024teamsllmagentsexploit} & Fang et al. & 2024-06-02 & 2 & 2025-03-30 & \\\hdashline
NYU CTF Dataset~\cite{shao2024nyuctfdatasetscalable} & Shao et al. & 2024-06-08 & 3 & 2025-02-18 & NeurIPS'24 (WS)\\\hdashline
PenHeal~\cite{huang2023penheal} & Hyuang et al. & 2024-07-25 & & & AutonomousCyber'24 (WS) \\\hdashline
Cybench~\cite{zhang2024cybench} & Zhang et al. & 2024-08-15 & 4 & 2025-04-12 & \\\hdashline
AutoPenBench~\cite{gioacchini2024autopenbenchbenchmarkinggenerativeagents} & Gioacchini et al. & 2024-10-04 & 2 & 2024-10-28 & \\\hdashline
Towards Automated Penetration Testing~\cite{isozaki2024towards} & Isozaki et al. & 2024-10-22 & 4 & 2025-02-21 & \\\hdashline
AutoPT~\cite{wu2024autoptfarend2endautomated} & Wu et al. & 2024-11-02 & & &  \\\hdashline
HackSynth~\cite{muzsai2024hacksynthllmagentevaluation} & Muzsai et al. & 2024-12-02 & & & \\\hdashline
Vulnbot~\cite{kong2025vulnbot} & Kong et al. & 2025-01-23 & & & \\\hdashline
On the Feasibility of Using LLMs to Execute Multistage Network Attacks~\cite{singer2025feasibility} & Singer et al. & 2025-01-27 & 3 & 2025-05-16 & \\\hdashline
Can LLMs Hack Enterprise Networks?~\cite{happe2025can} & Happe et al. & 2025-02-06 & & & \\\hdashline
RapidPen~\cite{nakatani2025rapidpenfullyautomatediptoshell} & Nakatani et al. & 2025-02-23 & & & \\
\bottomrule
\end{tabular}
\end{center}
\label{table:papers}
\end{table*}

\section{The Analyzed Publications}

This section gives a short overview about the included papers, the relationship between them, and their respective novelty. All included papers are detailed in Table~\ref{table:papers}.

\paragraph{Initial Papers} Two papers were referenced by all other reviewed papers: \textit{Getting pwn'd by AI}~\cite{happe2023getting} (July 2023) and \textit{PentestGPT}~\cite{deng2024pentestgpt} (August 2023). The former implemented a closed feedback-loop between an LLM and a target to autonomously perform a privilege-escalation attack; the latter incorporated LLMs with human instructions and feedback to interactively hack CTF machines. They also introduced the influential \textit{Pentest-Task-Tree} for strategy planning.

\paragraph{Autonomous Exploitation} Subsequent papers studied autonomous exploitation within different domains ranging from web-applications~\cite{fang2024llmagentsautonomouslyhack, wu2024autoptfarend2endautomated}, to post-breach attacks against Linux-~\cite{happe2024llms} or Windows-based~\cite{xu2024autoattacker} systems. LLMs have been used to autonomously exploit known one-day vulnerabilities~\cite{fang2024llmagentsautonomouslyexploit, nakatani2025rapidpenfullyautomatediptoshell} or even zero-day vulnerabilities~\cite{fang2024teamsllmagentsexploit}. PenHeal~\cite{huang2023penheal} introduced a dual-purpose prototype implementing both offensive as well as defensive capabilities.

\paragraph{State/Context Management} LLMs only have access to limited context to store their current state. Publications experimented with different ways of either reducing the used state~\cite{happe2024llms} or using RAG for both internal or external data storage~\cite{xu2024autoattacker, nakatani2025rapidpenfullyautomatediptoshell, huang2023penheal}. A \textit{planner} component for creating a high-level strategy was used by multiple publications~\cite{fang2024teamsllmagentsexploit, happe2025can, muzsai2024hacksynthllmagentevaluation, xu2024autoattacker, kong2025vulnbot}. Multiple papers were influenced by pentestGPT's \textit{Pentest-Task-Tree}~\cite{happe2025can, huang2023penheal}.

\paragraph{Benchmarks/Testbeds} Four papers focused on the introduction of a new penetration-testing benchmark and included their respective LLM-guided offensive prototype for benchmark evaluation. Two of them~\cite{shao2024nyuctfdatasetscalable, zhang2024cybench} were based on live real-world CTF events, while two were based on virtual CTF machines~\cite{isozaki2024towards, gioacchini2024autopenbenchbenchmarkinggenerativeagents}.

\paragraph{Targeting Networks} Recent publications switched their target from single-host targets to attacking whole organization networks~\cite{happe2025can, singer2025feasibility} spanning multiple computer systems.

\section{Results}

\begin{table*}[ht]
\caption{Testbed Overview. \textit{Testcases} can either be reused (R) from e.g. CTFs or CVEs, created from scratch (S) for the benchmark, or reused from another benchmark (B). The \textit{Implementation (Impl.)} can be based upon Container (C) or Virtual Machines (VM). \textit{Provenance} is denoted as released (R) if the benchmark is publicly released, documented (D) if it is not released but enough information, e.g., CVEs, are provided to reproduce the benchmark, and coarse (C) if only rough categories and not concrete vulnerabilities are given. \textit{\# Tasks} counts distinct test-cases which can include a number of vulnerabilities (\textit{\# Vuln.}). If achieving a task includes multiple vulnerabilities but the overall number of vulnerabilities is not stated, a ``?'' is used. \textit{Subtasks} indicates if a task is deconstructed into individual sub-tasks which are tracked. \textit{Linux}/\textit{Windows}/\textit{Web}/\textit{Other} denotes the target domain. \textit{Target} describes the target implementation: \textit{localhost} indicates that attacker and target run on the same computer, \textit{single-host} that the attacker targets a single network computer, and \textit{network} that a whole network-range is the attacker's target.}
\begin{center}
\begin{tabular}{lllllllllllll}
\toprule
Publication & \rot{Testcases} & \rot{Impl.} & \rot{Provenance} & \rot{Sources} & \rot{\# Tasks} & \rot{Subtasks} & \rot{\shortstack[l]{\# Vuln.}} & \rot{Linux} & \rot{Windows} & \rot{Web} & \rot{Other} & \rot{Target}\\
\midrule
Getting pwned by AI~\cite{happe2023getting} & R & VM & R & lin.security & 1 & & ? & \ok{} & & & & localhost \\\hdashline
LLMs as Hackers~\cite{happe2024llms} & S & VM & R & THM & 12 & \ok{} & 12 & \ok{} & & & & localhost \\\hdashline
Autonomously Hack Websites~\cite{fang2024llmagentsautonomouslyhack} & S & & C & & 15 & & 15 & & & \ok{} & & single-host\\\hdashline
Autonomously Exploit One-day Vulns.~\cite{fang2024llmagentsautonomouslyexploit} & S & & D & CVEs & 15 & & 15 & \ok{} & & \ok{} & \ok{} & single-host\\\hdashline
Exploit Zero-Day Vulnerabilities~\cite{fang2024teamsllmagentsexploit} & S &  & D & CVEs & 15 & & 15 & & & \ok{} & & single-host\\\hdashline
PenHeal~\cite{huang2023penheal} & R & VM & R & metasploitable & 1 & & 10 & \ok{} & & \ok{} & & single-host\\\hdashline
AUTOPENBENCH~\cite{gioacchini2024autopenbenchbenchmarkinggenerativeagents} & S & C & R & basic + CVEs & 33 & \ok{} & 33 & \ok{} & & \ok{} & \ok{} & single-host \\\hdashline
HackSynth~\cite{muzsai2024hacksynthllmagentevaluation} & R & & R & picoCTF, OTW & 200 & & 200 & \ok{} & & \ok{} & \ok{} & single-host\\\hdashline
Vulnbot~\cite{kong2025vulnbot} & B & & - &  \cite{gioacchini2024autopenbenchbenchmarkinggenerativeagents, isozaki2024towards} & & & & & & & & single-host\\\hdashline
Multistage Network Attacks~\cite{singer2025feasibility} & S & & R & VulnHub & 13 & \ok{} &  152 & \ok{} & & & & network\\\hdashline
pentestGPT~\cite{deng2024pentestgpt} & R & VM & R & HTB, VulnHub & 13 & \ok{} & 182 & \ok{} & \ok{} & \ok{} & & single-host\\\hdashline
Can LLMs hack Enterprise Networks?~\cite{happe2025can} & R & VM & R & GOAD & 15+ & \ok{} &  ? & & \ok{} & & & network\\\hdashline
Towards Automated Penetration Testing~\cite{isozaki2024towards} & S & VM & R & VulnHub & 13 & &  162 & \ok{} & & & & single-host \\\hdashline
AutoAttacker~\cite{xu2024autoattacker} & S & VM & C & & 14 & & 14 & \ok{} & \ok{} & & & single-host \\\hdashline
CyBench~\cite{zhang2024cybench} & S & C & R & CTFs & 40 & \ok{} & & \ok{} & & \ok{} & \ok{} & single-host\\\hdashline
NYU CTF Dataset~\cite{shao2024empiricalevaluationllmssolving, shao2024nyuctfdatasetscalable} & S & C & R & CTFs & 26 & & & & & \ok{} & \ok{} & single-host\\\hdashline
RapidPen~\cite{nakatani2025rapidpenfullyautomatediptoshell} & R & VM & R & HTB & 1 & & & & \ok{} & & & single-host \\\hdashline
AutoPT~\cite{wu2024autoptfarend2endautomated} & R & VM & R & VulnHub & 17 & & 20 & & & \ok{} & & single-host \\
\bottomrule
\end{tabular}
\end{center}
\label{table:overview}
\end{table*}

The 19 analyzed papers leveraged 18 testbeds of which 7 created new benchmarks reusing existing CTF cases while 10 papers implement a new benchmark from scratch. A single paper (vulnbot~\cite{kong2025vulnbot}) reused two existing benchmarks for their evaluation.

\subsection{Testbed Design}

Table~\ref{table:overview} shows the overall design of the analyzed testbeds, detailing the benchmarks' target systems, its provenance, and implementation choices.

\paragraph{Target Systems} Testbeds commonly emulated Linux, Windows, or Web-based systems. Two benchmarks (Cybench~\cite{zhang2024cybench} and NYU~\cite{shao2024nyuctfdatasetscalable}) included traditional CTF challenges such as cryptography, forensics, reversing, and exploit-generation. All but two papers used singular hosts as their target systems, either by providing a direct shell connection or by designating the target by its singular IP network address. The remaining two benchmarks used simulated networks containing multiple virtual machines. One benchmark~\cite{xu2024autoattacker} created a test network, but the test-cases themselves were only targeting individual systems and thus were counted as a single-host benchmark.

\paragraph{Reproducibility} One benefit of reusing existing CTF tasks was improved reproducibility as the included tasks are typically publicly available---albeit sometimes behind a paywall. Of the self-built benchmarks, only a single one~\cite{happe2024llms} was publicly available. Of the remaining five benchmarks, two were specified through their implemented CVEs and thus reproducible. Finally, three benchmarks only provided coarse documentation, e.g., used attack classes, thus limiting their reproducibility.

\paragraph{Tasks} Benchmarks contained between 1--200 high-level tasks (average: $26.1$, median: $15$), typically provided through a separate virtual machine or container. One benchmark---the NYU CTF dataset~\cite{shao2024nyuctfdatasetscalable}---contained 200 tasks but only few penetration-testing specific cases (19 web pen-testing tasks). Depending on the used benchmark, high-level tasks were separated into multiple steps, subtasks, or vulnerabilities. There was no common vocabulary nor semantics for what constitutes a sub-tasks.

\paragraph{Sub-Tasks} All reviewed papers provided a binary success rate: a test-case is either completed successfully or not. 6 publications provided fine-grained sub-task analysis. They differed in how they identified and mapped the needed sub-steps.

Happe et al.~\cite{happe2024llms} performed an analysis of captured log traces utilizing human pen-testers to match executed sub-tasks to MITRE ATT\&CK tactics and procedures. Deng et al.~\cite{deng2024pentestgpt} used NIST 800-115 to classify tasks into 10 broad categories and showed how testing trajectories traverse through these categories. Other papers create an a-priori list of tasks that must be executed by an attacker to achieve exploitation. These steps were often created manually---by the authors or dedicated pen-testers---or by analyzing publicly available CTF walk-throughs~\cite{isozaki2024towards}. AutoPenBench~\cite{gioacchini2024autopenbenchbenchmarkinggenerativeagents} defined both ``gold steps'' as well as milestones. Milestones are either defined by executing specific commands stated within the golden steps or by achieving tasks. LLMs are employed to match log traces against the golden steps and milestones, and human quality control is additionally performed.

\subsection{Experiment Design}

\begin{table*}[ht]
\caption{LLM used within Publications. \textit{Used Models} gives the list of used LLMs as stated within the respective publication. \textit{Walled-Garden} indicates that at least a LLM without access to the model weights (typically cloud-hosted) was used. \textit{Open-Weight} indicates that at least a LLM with publicly available model weights was used. SLM (``Small Language Model'') indicates inclusion of a LLM with less than 16b parameters which implies usability on consumer-grade graphics cards. Finally, \textit{Reasoning Model} indicates that at least one reasoning model was used during the benchmark.}
\begin{center}
\begin{tabular}{lcccccp{8cm}}
\toprule
 Publication & \rot{Walled-Garden} & \rot{Open-Weight} & \rot{SLM} & \rot{Reasoning Model} & \rot{\# of LLMs evaluated } & Used LLMs\\
 \midrule
 Getting pwned by AI~\cite{happe2023getting}    & \ok{} & & & & 1 & gpt-3.5-turbo \\\hdashline
 LLMs as Hackers~\cite{happe2024llms}           & \ok{} & \ok{} & \ok{} & & 4 & gpt-3.5-turbo, gpt-4-turbo, llama3:70b, llama3:8b\\\hdashline
 Autonomously Hack Websites~\cite{fang2024llmagentsautonomouslyhack} & \ok{} & \ok{} & \ok{} & & 10 & gpt-3.5, gpt-4, openhermes-2.5-mistral:7b, llama2-chat:70b, llama2-chat:13b, llama2-chat:7b, mixtral:8x7b, mistral-instruct-v2:7b, nous-hermes-2-yi:34b, openchat 3.5\\\hdashline
 Autonomously Exploit One-day Vulns.~\cite{fang2024llmagentsautonomouslyexploit} & \ok{} & \ok{} & \ok{} & & 10 & gpt-3.5, gpt-4, openhermes-2.5-mistral:7b, llama2-chat:70b, llama2-chat:13b, llama2-chat:7b, mixtral:8x7b, mistral-instruct-v2:7b, nous-hermes-2-yi:34b, openchat 3.5\\ \hdashline
 Exploit Zero-Day Vulnerabilities~\cite{fang2024teamsllmagentsexploit} & \ok{} & \ok{} & & & 3 & gpt-4-0125-preview, llama-3.1:405b, qwen-2.5:72b\\\hdashline
 PenHeal~\cite{huang2023penheal} & \ok{} & & & & 1 & gpt-4 \\\hdashline
AUTOPENBENCH~\cite{gioacchini2024autopenbenchbenchmarkinggenerativeagents} & \ok{} & & & & 1 & gpt-4o \\\hdashline
 HackSynth~\cite{muzsai2024hacksynthllmagentevaluation} & \ok{} & \ok{} & \ok{} & & 8 & gpt-4o, gpt-4o-mini, llama-3.1:8b, llama-3.1:70b, qwen2:72b, mixtral:8x72b, phi-3-mini-4k, phi-3.5-MoE \\\hdashline
 Vulnbot~\cite{kong2025vulnbot} & \ok{} & \ok{} & & & 3 & gpt-4o, llama3.3:70b, llama3.1:405b\\\hdashline
 Multistage Network Attacks~\cite{singer2025feasibility} & \ok{} & & & & 3 & gpt-4o, gemini 1.5 pro, sonnet 3.5\\\hdashline
 pentestGPT~\cite{deng2024pentestgpt} & \ok{} & & & & 3 & gpt-3.5, gpt-4, bard \\\hdashline
 Can LLMs hack Enterprise Networks?~\cite{happe2025can} & \ok{} & & & \ok{} & 2 & o1, gpt-4o \\\hdashline
 Towards automated penetration testing~\cite{isozaki2024towards} & \ok{} & \ok{} & & & 2 & gpt-4o, llama3.1:405b\\\hdashline
 AutoAttacker~\cite{xu2024autoattacker} & \ok{} & \ok{} & \ok{} & & 4 & gpt-3.5, gpt-4, llama2-chat:7b, llama2-chat:70b\\\hdashline
 CyBench~\cite{zhang2024cybench} & \ok{} & \ok{} & & \ok{} & 8 & GPT-4o, o1-preview, Claude 3 Opus, Claude 3.5 Sonnet, Gemini 1.5 Pro, Mixtral-Instruct:8x22b, Llama-3-Chat:70B, Llama-3.1-Instruct:405B\\\hdashline
 NYU CTF Dataset\cite{shao2024empiricalevaluationllmssolving, shao2024nyuctfdatasetscalable} & & \ok{} & &  & 5 & Mixtral-Instruct-v0.1:8x7B, deepseek-coder-instruct:33b, llama3-instruct:70b, wizardlm2:8x22b, Llama-3-Instruct:70B\\\hdashline
 RapidPen~\cite{nakatani2025rapidpenfullyautomatediptoshell} & \ok{} & & & & 1 & gpt-4o \\\hdashline
AutoPT~\cite{wu2024autoptfarend2endautomated} & \ok{} & & & & 3 & gpt-3.5, gpt-4, gpt-4-mini \\
\bottomrule
\end{tabular}
\end{center}
\label{table:llm_selection}
\end{table*}

\paragraph{LLM Selection} Experiments within the reviewed papers typically analyzed between 1--10 LLMs (average $4.0$, median $3$). Two papers~\cite{gioacchini2024autopenbenchbenchmarkinggenerativeagents, wu2024autoptfarend2endautomated} were performing preliminary experiments to identify and remove LLMs with insufficient capabilities from their evaluation set. An overview of the used LLMs is given in Table~\ref{table:llm_selection}.

The prevalently used LLM-family were OpenAI's non-reasoning models which were included in all but one publication. Distant second was Meta's Llama family which was included in roughly half of the papers while Mistral's LLM-family being included in a quarter of the papers.

There is no authoritative definition for Small Language Models (SLMs), but given the common understanding that these are models that are able to run on edge-devices such as desktops, we count all LLMs using less that 16b parameters as SLMs. Using 4bit quantization, these models fit into 8 gigabyte of memory while being able to generate adequate token counts for interactive use on desktop-class computers. A quarter of the publications (5 papers) included SLMs within their evaluation. Reasoning LLMs, which are a recent addition to LLM capabilities, were included in two of the reviewed publications.

\paragraph{Sample Size} Table~\ref{table:measures} details sample sizes and upper-bounds encountered within reviewed publications. During experiments, between 1--6 test-runs/samples were executed per LLM (average $4.6$, median $5$). Only half of the papers detailed the length of the captured samples. If reported, the maximum sample duration was either defined through an upper-bound of executed steps/commands (15--60 steps, on average 30 steps), or through introducing a maximum sample duration (ranging from 10 minutes to 48 hours). Two papers supplemented test-cases in addition to their defined benchmark. \textit{PentestGPT}~\cite{deng2024pentestgpt} used additional CTF test-cases, while Fang et al.~\cite{fang2024llmagentsautonomouslyhack} targeted 50 additional hand-curated web-sites of undefined provenance.

\paragraph{Baselines} Papers offered baselines to compare their LLM-driven prototypes against. Human baselines were either provided through quantitative analysis of log traces produced by human penetration testers~\cite{happe2024llms} or through analysis of human-generated example walk-throughs (see Section~\ref{subtasks}). Automated baselines were created by running traditional automated security security tooling (e.g., ZAP\footnote{Zed Attack Proxy, \url{https://www.zaproxy.org/}} or metasploit\footnote{\url{https://www.metasploit.com/}}) or existing LLM-driven prototypes (2 papers used prototypes from the respective authors prior work, while 5 papers used \textit{pentestGPT}~\cite{deng2024pentestgpt}). Table~\ref{table:measures} gives an overview of the baselines used within reviewed publications.

\subsection{Measures and Analysis}

All of the reviewed papers tracked the success rates of their prototypes, typically split-up per test-case and/or per tested LLM. Complex and realistic vulnerabilities often consist of multiple causally-dependent tasks, e.g., an autonomous agent must initially enumerate the system, identify a vulnerability, and subsequently exploit it; only tracking the binary outcome cannot detail LLMs’ capabilities with those intermediate steps. 6 out of the 18 analyzed prototypes tracked these mentioned sub-steps.

Half of the papers (8) captured input/output token counts and used them for cost estimates, typically stated in US\$. This is a convenient estimate of a prototype's efficiency as occurring costs are highly dependent upon the used LLM and their tokenizers. LLMs commonly have asymmetric pricing for input and output tokens; their pricing frequently changes over time. Due to this dynamic pricing regime, stating the occurred costs allows for easier long-term comparison of the prototype's efficiency.

Detailed information about executed commands was sparse. 9 papers tracked the amount of executed system commands, either directly or indirectly through their stated ``round'' number. Of these, roughly half (4 papers) classified executed commands or provided a list of frequently executed commands. 7 papers additionally tracked the amount of invalid commands and further detailed why command execution resulted in errors.

Every paper performed a qualitative analysis of error traces, ubiquitously by the respective authors.

\section{Discussion and Recommendations}

\subsection{Technology/Implementation Choices}
\label{technology_choices}

All testbeds were implemented using either containers or virtual machines (VMs). The chosen virtualization technology impacts testbed design, i.e., using containers effectively prevents using Windows-based test-cases. Containers and VMs also provide different security boundaries which impact the testbed's safety, e.g., containers cannot be used to safety provide kernel-level vulnerabilities.

Testbeds were often intertwined with an agent prototype or framework. While this does not enforce the use of the respective agent framework, it might ease the integration of a potenial attack prototype into the target testbed.

Using commercial cloud-based CTF VMs, e.g., HackTheBox or TryHackMe, has implications on availability and reproducibility. Cloud providers do not guarantee testbed stability, e.g., used software versions. As Isozaki et al.~\cite{isozaki2024towards} noted, retired HackTheBox machines are only available for premium accounts. Commercial offerings typically do not detail their setup nor provide build-instructions for provided CTF challenges.

\paragraph{Recommendation} Evaluate technology choices esp. for safety and security implications, e.g., if system-level attacks are part of the testbed, virtual machines should be used. The testbed or building instructions should be available ``offline'' to allow for reproducibility.

\subsection{Benchmark Composition}
\label{composition}

\begin{table*}[ht]
\caption{The experiment design utilized within the reviewed publications. If supplemental test-cases were used in addition to the described testbed, they are noted as \textit{Additional Test-Cases}. We detail the number of used LLMs (\textit{\# LLMs}) as well as the number of independent test-runs/samples (\textit{Sample Size}). We also document the stop condition of the performed experiment, which was either given as a maximum number of steps (\textit{Max. Steps/Sample}) or a maximum time (in Minutes) per Sample (\textit{Max. Time/Sample}).}

\begin{center}
\begin{tabular}{llrrrrr}
\toprule
 Publication & \shortstack[l]{Additional\\ Test-Cases} & \# LLMs & Sample Size & \shortstack[r]{Max.\\ Steps/Sample} & \shortstack[r]{Max.\\ Time/Sample}\\
 \midrule
 Getting pwned by AI~\cite{happe2023getting} & & 1 & &\\\hdashline
 LLMs as Hackers~\cite{happe2024llms} & & 4 & 1 & 60\\\hdashline
 Autonomously Hack Websites~\cite{fang2024llmagentsautonomouslyhack} & 50 web sites & 10 & 5 & & 10\\\hdashline
 Autonomously Exploit One-day Vulns.~\cite{fang2024llmagentsautonomouslyexploit} & & 10 & 5 & \\\hdashline
 Exploit Zero-Day Vulnerabilities~\cite{fang2024teamsllmagentsexploit} & & 3 & 5 & &\\\hdashline
 PenHeal~\cite{huang2023penheal} & & 1 & 3 & \\\hdashline
AUTOPENBENCH~\cite{gioacchini2024autopenbenchbenchmarkinggenerativeagents} & & 1 & 5& 30/60 & \\\hdashline
 HackSynth~\cite{muzsai2024hacksynthllmagentevaluation} & & 8 & & 20 \\\hdashline
 Vulnbot~\cite{kong2025vulnbot} & & 3 & 5 & 15/24\\\hdashline
 Multistage Network Attacks~\cite{singer2025feasibility} & & 3 & 5\\\hdashline
 pentestGPT~\cite{deng2024pentestgpt} & picoCTF, HTB & 3 & \\\hdashline
 Can LLMs hack Enterprise Networks?~\cite{happe2025can} & & 2 & 6 & & 120 \\\hdashline
 Towards automated penetration testing~\cite{isozaki2024towards} & & 2 & 1\\\hdashline
 AutoAttacker~\cite{xu2024autoattacker} & & 4 & 3\\\hdashline
 CyBench~\cite{zhang2024cybench} & & 8 & & 15\\\hdashline
 NYU CTF Dataset\cite{shao2024empiricalevaluationllmssolving, shao2024nyuctfdatasetscalable} & & 5 & 5 & & 2880\\\hdashline
 RapidPen~\cite{nakatani2025rapidpenfullyautomatediptoshell} & & 1 & 10 & & \\\hdashline
AutoPT~\cite{wu2024autoptfarend2endautomated} & & 3 & 5 & 15 & \\
\bottomrule
\end{tabular}
\end{center}
\label{table:experiment_design}
\end{table*}

A benchmark’s task composition is of utmost importance for its construct validity, i.e., how well the benchmark approximates real-life security practitioners’ work and challenges.

Benchmark tasks were typically mapped to existing attack vector classification schemes such as MITRE ATT\&CK or the OWASP Top 10 Web Vulnerabilities. A reverse mapping, i.e., showing the coverage that a benchmark provides of a hacking discipline, was not provided. A potential reason for this is that while classification schemes for attack vectors exist within penetration testing, they do not provide a hacking methodology and thus cannot be used to structure penetration-tests.

Not having an authoritative source of attack vectors opens up task composition for discussion. For example, should basic file operations (reading, writing, or uploading files) or navigation within the target system, be part of a security benchmark? AutoAttacker~\cite{xu2024autoattacker} and HackSynth~\cite{muzsai2024hacksynthllmagentevaluation} contain tasks that verify that LLMs are able to perform these basic system operations. Fang et al.~\cite{fang2024llmagentsautonomouslyexploit, fang2024teamsllmagentsexploit} call existing benchmarks ``toy problems'' and create their own benchmark based upon CVEs, i.e., software with known vulnerabilities. While they never define the term ``toy problems'', it could be explained by benchmarks including the mentioned basic tasks such as file operations. On the other hand, benchmarks such as NYU~\cite{shao2024nyuctfdatasetscalable} or CyBench~\cite{zhang2024cybench} are themselves partially based on CVEs, thus while often called ``toy benchmarks'', they are comparable with a Fang's created benchmark.

Another issue arises from using virtual machines that are originally intended for penetration-tester education, such as \textit{lin.security}, \textit{metasploitable2}, or \textit{GOAD}. While they offer the benefit of matching penetration-tester real-life experiences, they typically contain multiple parallel vulnerabilities within the same virtual machine and their included attack vectors are often insufficiently documented. For example, Happe et al.~\cite{happe2023getting} initially used the \textit{lin.bench} virtual machine for evaluating Linux privilege escalation techniques. In later works~\cite{happe2024llms}, they switched to a bespoke benchmark consisting of a single VM per vulnerability class as LLMs otherwise would always exploit the same ``simple'' attack paths within lin.security. PenHeal~\cite{huang2023penheal} uses a single metasploitable2 virtual machine as a testbed and details the included attack classes within their paper. Concurrent walk-throughs\footnote{\url{https://docs.rapid7.com/metasploit/metasploitable-2-exploitability-guide}} indicate that additional attack classes are included within metasploitable2, thus invalidating coverage metrics. Similarly, Happe et al.~\cite{happe2025can} utilize GOAD as an Active Directory testbed containing 5 windows server VMs and 30 Active Directory users. There is no authoritative documentation detailing all vulnerabilities and attack paths within this testbed. Partial documentation\footnote{\url{https://orange-cyberdefense.github.io/GOAD/img/diagram-GOAD_compromission_Path_dark.png}} indicates the existence of dozens of potential attack paths which often have to be combined to enable further exploitation. Given this situation, the evaluation can only count the amount of compromised systems and users, but cannot give an estimate of achieved vulnerability coverage.

\paragraph{Recommendation} Ground the test-cases in reality by using ``Top 10 lists'' for broad guidance, but provide detailed information which attack vectors were included within the testbed.

\subsection{Practitioners’ Work: Security vs. Pen-Testing Challenges}
\label{practitioners_work}

Testbeds based upon CTF -Challenges~\cite{zhang2024cybench, muzsai2024hacksynthllmagentevaluation, shao2024nyuctfdatasetscalable} contain attack vectors belonging to broad categories such as reversing, forensics or exploitation-writing challenges in addition to typical penetration-testing activities such as web exploitation. Recent empirical research~\cite{happe2023understanding} into penetration testers’ tasks indicates a split between people working within the field of security: security researchers and security practitioners. For the former, challenges such as reversing or exploit generation are highly relevant, while for the latter, finding security misconfigurations or exploiting known vulnerabilities is more relevant. Penetration-Testers in the field typically fall into the security practitioner category. In addition, forensics is typically delegated to dedicated personnel that are not performing penetration testing. While CTF-based challenges mirror the security field as a whole, they might not provide a good proxy for penetration testing practices.

Another mismatch are Assumed Breach scenarios, which are commonly performed by security practitioners. In these scenarios, the attacker is already situated within the target environment and performs network-based attacks. They commonly have to combine singular low-level vulnerabilities into vulnerability chains to breach their targets. While CTF challenges' atomic exercises simulate exploiting those low-level vulnerabilities, they often do not include those multi-step attack chains or limit the included attack-chains to a single target machine. In contrast, more network-oriented benchmarks (\cite{happe2025can,singer2025feasibility}) typically include multi-step scenarios spanning multiple virtual machines.

All reviewed benchmarks were Jeopardy-style CTFs. Attacker / Defender style benchmarks would provide additional realism by including dynamism into the testbed, e.g., configuration changes, active adversaries, stealthiness, detection engineering, and both implementing and dealing with countermeasures. Of the reviewed testbeds, the network-based testbeds~\cite{happe2025can,singer2025feasibility} would be best suited for extending into Attacker/Defender style testbeds.

\paragraph{Recommendation} Analyze your audience and design your test-cases accordingly.

\subsection{Training Data Contamination}
\label{training_data_contamination}

\begin{table*}[ht]
\caption{Measures used within publications. We differentiate between baselines used: \textit{Human Baselines} using humans to perform the test-cases, \textit{LLM-Prototype} used existing LLM-based prototypes to gather a baseline while \textit{Trad. Tooling} used existing automated penetration-testing tooling for generating a baseline. We note the tools used, which were typically the \textit{ZAP Attack Proxy} (Z) or \textit{Metasploit} (M). We note if the the benchmarks capture binary \textit{Success Rates} (``test-case successful yes/no'') or offer fine-grained \textit{Progression Rates} (``test-case 60\% completed''). The columns \textit{Tokens} and \textit{Costs} denote if the resulting test-run costs were given as token counts or as amount of US\$. We note if metrics include information about the number of successfully executed commands as well as the amount of errors occurred during command execution. Finally, we indicate if the respective publication provided a classification of executed commands or encountered errors.}
\begin{center}
\begin{tabular}{l|lll|ll|ll|llll}
\toprule
 Publication & \rot{Human Baseline} & \rot{LLM-Prototype} & \rot{Trad. Tooling} & \rot{Success Rate} & \rot{Progression Rate} & \rot{Tokens}& \rot{Costs} & \rot{Command Count} & \rot{\shortstack[l]{Invalid\\ Command Count}} & \rot{\shortstack[l]{Command\\ Classification}} & \rot{Error Classification}\\
 \midrule
 Getting pwned by AI~\cite{happe2023getting} & & & & \ok{} & & & & & & \\\hdashline
 LLMs as Hackers~\cite{happe2024llms} & \ok{} & \ok{} & & \ok{} & \ok{} & \ok{} & \ok{} & \ok{} & & & \\\hdashline
 Autonomously Hack Websites~\cite{fang2024llmagentsautonomouslyhack} & & & & \ok{} & & & \ok{} & \ok{} & & \\\hdashline
 Autonomously Exploit One-day Vulns.~\cite{fang2024llmagentsautonomouslyexploit} & & & Z, M & \ok{} & & \ok{} & \ok{} & \ok{} & & & \\\hdashline
 Exploit Zero-Day Vulnerabilities~\cite{fang2024teamsllmagentsexploit} & & \ok{} & Z, M& \ok{} &  & \ok{} & \ok{} & & & & \\\hdashline
 PenHeal~\cite{huang2023penheal} & & \ok{} & & \ok{} & & & & \ok{} & & \\\hdashline
AUTOPENBENCH~\cite{gioacchini2024autopenbenchbenchmarkinggenerativeagents} & & & & \ok{} & \ok{} & & & & \ok{} & & \ok{} \\\hdashline
 HackSynth~\cite{muzsai2024hacksynthllmagentevaluation} & & \ok{} & & \ok{} & & \ok{} & \ok{} & \ok{} & & \ok{} & \\\hdashline
 Vulnbot~\cite{kong2025vulnbot} & & \ok{} & & \ok{} & & & & & \ok{} & & \ok{} \\\hdashline
 Multistage Network Attacks~\cite{singer2025feasibility} & & \ok{} & & \ok{} & \ok{} & & & \ok{}& \ok{} & & \\\hdashline
 pentestGPT~\cite{deng2024pentestgpt} & & & & \ok{} & \ok{} & & \ok{} & \ok{} & \ok{} & \ok{} & \ok{} \\\hdashline
 Can LLMs hack Enterprise Networks?~\cite{happe2025can} & & & & \ok{} & \ok{} & \ok{} & \ok{} & \ok{} & \ok{} & \ok{} & \ok{} \\\hdashline
 Towards automated penetration testing~\cite{isozaki2024towards} & & \ok{} & & \ok{} & & & & \ok{} & \ok{} & \ok{} & \ok{} \\\hdashline
 AutoAttacker~\cite{xu2024autoattacker} & & & & \ok{} & & & \ok{} & & & & \\\hdashline
 CyBench~\cite{zhang2024cybench} & & & & \ok{} & \ok{} & & & & & & \\\hdashline
 NYU CTF Dataset\cite{shao2024empiricalevaluationllmssolving, shao2024nyuctfdatasetscalable} & & & & \ok{} & & & & & \ok{} & & \ok{} \\\hdashline
 RapidPen~\cite{nakatani2025rapidpenfullyautomatediptoshell} & & & & \ok{} & & & \ok{} & & & & \ok{} \\\hdashline
AutoPT~\cite{wu2024autoptfarend2endautomated} & & & & \ok{} & & & \ok{} & & & & \ok{}\\
 \bottomrule
\end{tabular}
\end{center}
\label{table:measures}
\end{table*}

Publicly available testbeds will be included within LLM training data eventually.  To prevent overfitting, multiple publications selected vulnerabilities that have a CVE publication date that is after the tested LLM’s training cut-off date. This assumes that there is no research or exploit released prior to the publication of a CVE. This is---by definition---not the case for \textit{0days}, i.e., vulnerabilities that are actively exploited before a remediation is provided by defenders within their public announcement as part of coordination disclosure procedures. In addition, using a cut-off date prevents inclusion of relevant older techniques in the benchmark, which is especially important in scenarios that emulate common real network vulnerabilities as corporate networks often contain legacy protocols or services. A potential solution would be to make all identifiers within a benchmark parametrizable or randomized. This would allow each benchmark instantiation to contain unique usernames, hostnames, passwords, or file paths. In addition, benchmarks and their documentation should contain canaries that allow better detection if a benchmark is included within an LLM's training data.

Another issue is Goddhart’s law: ``\emph{when a measure becomes a target, it ceases to be a good measure}''~\cite{chrystal2003goodhart}. In the security domain this is also related to the Red Queen's race~\cite{blank2017red} as we have have active adversaries. Every time a new Top 10 list of vulnerabilities is published and defenders implement countermeasures for the respective Top 10 items, attackers switch to additional attack vectors, i.e., the attacks that just did not make it within the Top 10s. As these attacks now rise in prominence, the subsequent list of Top 10 items will contain those abused vulnerabilities and attackers again will switch to the items that are just outside of the Top 10. Using historic training data thus might teach an LLM attack vectors that are currently ``out-of-style''.

\paragraph{Recommendation} Make identifiers within the testbed parametrizable and include canaries in both your testbed and its documentation.

\subsection{Reproducibility of Baselines}
\label{baselines}

Human baselines are inherently not reproducible. In addition, automated tooling that depends upon human interactions, e.g., using pentestGPT as a baseline, can incorporate this human randomness in addition to the tooling-inherent randomness. Using LLM-guided baselines introduces problems with reproducibility due to their stochastic nature.

When using automation, choosing the right tooling is important: ZAP is a web vulnerability scanner and should only be used for benchmarks that consist primarily of web vulnerabilities. When used as a baseline, the utilized configuration should be documented. For example, if ZAP is used in its autonomous \textit{baseline scan mode}, by default, execution is stopped after one minute, which does not provide sufficient test coverage. In addition, ZAP is highly dependent upon its configured plugins and rule-sets, without stating those explicitly, the generated baselines are not reproducible.

\paragraph{Recommendation} Implement a baseline and include it into your testbed documentation. If the baseline is created through automated tooling include enough configuration data to make the baseline reproducible.

\subsection{Clean Test-Cases vs. Messy Real-Life}
\label{messy_life}

Test-cases should be reproducible, i.e., subsequent executions should provide stable results. This leads towards synthetic single-step atomic test-cases that can be evaluated individually. While being beneficial for benchmarking, this can negatively impact construct validity. Sommer and Paxson note in ``Outside the Closed World: On Using Machine Learning for Network Intrusion Detection''~\cite{sommer2010outside}, that synthetic testing environments can lead to an oversimplified understanding of adversarial behavior as they typically fail to capture the dynamic complexity and nuanced behaviors inherent in real-world systems and networks.

\paragraph{Multi-Step Attack Chains} Common attacks in real-world networks consist of multiple causally-dependent steps. For example, a user account must first be compromised using a network attack before it is subsequently used to execute a sensitive operation. Real-Life penetration-tests can commonly be described as complex attack graphs where nodes indicate states and edges offensive steps. Both, nodes being dependent upon multiple successful prior exploitation steps, as well as redundant parallel steps leading to the same compromised node, are common occurrences. To allow for better analysis of singular attack steps, testbeds often simplify multi-step attack chains into separate tests containing only a single attack step. This does not allow to test for end-to-end security exploitation, esp. does not allow testing LLMs for high-level strategizing and planning.

\paragraph{Exploits are often non-deterministic} Synthetic testbeds often assume attack steps to be deterministic. Real-life exploits are often stochastic, e.g., when exploiting the well-known \textit{EternalBlue}\footnote{\url{https://de.wikipedia.org/wiki/EternalBlue}} Windows vulnerability, the probability of a successful compromise is inherently low and subject to variability; in some cases, executing the exploit can crash the target system. Synthetic testbeds, by their nature, often ignore these probabilistic effects and the cascading consequences of exploit-induced system instability.

\paragraph{Side-Effects in the Real-World} Many security-relevant operations can evoke side-effects, e.g., if a non-deterministic exploit crashes the target system, the same system can subsequently not be used within an attack-graph. Another example is the existence of an active adversary, e.g., endpoint detection and response (EDR) software place within network testbeds. Synthetic testbeds often do not accurately model the consequence of such minute variations, and if these active countermeasures are disabled to accommodate the simulation, the intrinsic realism of the scenario is compromised. Without this dynamic interplay, synthetic benchmarks risk misrepresenting the true performance of automated attack strategies.

\paragraph{Background noise and activity} Another essential aspect of real-world networks is the presence of concurrent background activities that can be part of an attack graph. For example, periodic tasks are often executed within server systems, or users periodically interact in insecure ways with network services. These temporal patterns are critical when evaluating network-level attacks, e.g., \textit{pass-the-token} or \textit{pass-the-hash} attacks in enterprise networks. In a synthetic benchmark, these time-based nuances are typically flattened or entirely absent further distorting their real-world applicability.

\paragraph{Recommendation} Emulate real-life problems even if they are messy. Implement multi-step attack chains that emulate common attack trees.

\begin{table*}[ht]
\caption{Summary of our Recommendations}

\begin{center}
\begin{tabular}{lll}
\toprule
\textbf{Chapter} & \textbf{Recommendation} \\
 \midrule
 \ref{technology_choices}: Technology Choices & Evaluate technology choices esp. for safety and security implications. \\
 \ref{composition}: Benchmark Composition & Ground the benchmark in reality and provide information about included vulnerabilities. \\
 \ref{practitioners_work}: Practitioners' Work & Consider your audience and create relevant test-cases.\\
 \ref{training_data_contamination}: Training Data Contamination & Randomize identifier and include Canaries. \\
 \ref{baselines}: Baselines         & Provide baselines derived from humans or automated tooling (include configuration). \\
 \ref{messy_life}: Clean Test-Cases vs. Messy Life & Emulate real-life problems. \\
 \ref{subtasks}: Tracking Sub-Tasks & Use Sub-Tasks for fine-grained analysis and allow for automated task completion detection. \\
 \midrule
 \ref{llm_selection}: LLM Selection & Run at least one SotA LLM, one open-weight LLM, and, if feasible a SLM. \\
                                    & If feasible, use at least one OpenAI LLM to allow for comparison with existing research.\\
                                    & State your LLM's requirements and detail their configuration, e.g., temperature.\\
 \ref{experiment_design}: Experiment Design & Run at least 5 samples and set the limit of steps per sample to at least 32.\\
                                    & If provided, use baselines for comparison.\\
\ref{metrics_and_analytics}: Metrics and Analysis & Measure success rates, token utilization and occurred costs.\\
                                                  & Overview executed commands and their errors. \\
                                                  & Perform qualitative analysis of trajectories and include your methodology.\\
 \bottomrule
\end{tabular}
\end{center}
\label{table:recommendations}
\end{table*}

\subsection{Progress Tracking through Sub-Tasks}
\label{subtasks}

Command- or milestone-based progress tracking implicitly assumes that progress within penetration-testing can be linearized. Modern attack methodologies are moving away from waterfall-like models towards iterative approaches~\cite{7836624}. Due to their complex interactions, real-life attacks are often visualized through attack-trees~\cite{schneier1999attack} and attack graphs~\cite{LALLIE2020100219}, which incorporate parallel execution and dependencies between attack stages.

When using golden steps, an implicit assumption is that commands and tools used during penetration-testing are known before the experiment occurs as they need to be stated within the golden steps. This assumption might not hold, i.e., attack tools evolve over time, and newer LLMs learn those new tools through their training data. This can become problematic, e.g., if a golden step refers to the \textit{cme} command while the LLM uses its newer \textit{nxc} version, it might not be detected as successful sub-task completion.

CyBench~\cite{zhang2024cybench} provides an optional subtask tracking mode, called \textit{subtask-guided performance}. For each task a list of questions is defined, e.g., ``\emph{which files contain the account credentials}''? During execution, the attack prototype and its included LLM are tasked with answering the current relevant question. If it provides a correct answer, the attack prototype is assumed to have progressed to the next sub-task and is subsequently asked with the next relevant question. This is an implicit guidance mechanism and inherently alters the analyzed model’s performance.

\paragraph{Recommendation} We encourage using sub-tasks to allow for fine-grained analysis of traces. If you implement sub-tasks, devise means of automatic detection if a subtask has been achieved. We recommend to define sub-tasks through their expected result and not through invoked tool-calls. Detail which preconditions must be fulfilled to make execution of a sub-task viable, as well as which other sub-tasks become viable after a subtask has been achieved. We suggest showing the potential interactions between subtasks, e.g., through use of flow diagrams.

\subsection{LLM Selection}
\label{llm_selection}

Given the prevalence of OpenAI's non-reasoning LLMs within our reviewed publications, using a model out of this family allows for easier comparison to existing results. Of the open-weight models, the high-usage of Llama and Mistral models might be related to their easy availability (the initial Llama model was released in February 2023). Newer models, such as the DeepSeek- or qwen-family of models, are beginning to be used by researchers.

The small amount of used reasoning-models within reviewed publications could be related to their relative novelty. OpenAI's \textit{o1-preview} was announced in September 2023, out of the 19 publications, only 5 were initially published after the announcement. Of these, one paper was already using reasoning models while an older publication~\cite{zhang2024cybench} was updated to include reasoning models.

\paragraph{Recommendation} Run at least one State-of-the-Art LLM and a locally-run Small Language Model (SLM) which, by definition, will also be an open-weight LLM. The state-of-the-art LLM will typically be cloud-hosted. If suitable, use a model from the OpenAI family to allow for easy comparison with other publications. If an SLM is not reasonable for the task, use at least one open-weight LLM for testing. State your LLM requirements with regard to reasoning-capabilities, tool/function calling, structured output, and minimally required context size. State your chosen LLM configuration, e.g., the configured LLM temperature.

\subsection{Experiment Run Configuration}
\label{experiment_design}

Based on the averages gathered from the reviewed publications, we recommend performing at least 5 test-runs/samples per LLM with a maximum that's at least 32 steps. We encourage creating both human and automated baselines and recommend including extensive configuration information when automated tools are used for their creation (Section~\ref{baselines}).

\subsection{Gathered Metrics and Analysis}
\label{metrics_and_analytics}

All reviewed publications measured their performance through binary success rates. A third of the publications used sub-tasks to track the evaluated LLM's progress through their respective tasks. 10 out of the 18 evaluations used the occurring \$-cost to track their prototype's efficiency, half of them additionally noted their respective token utilization. All papers perform qualitative and quantitative analysis, typically a form of thematic analysis.

\paragraph{Recommendation} Include metrics for per-model and per-testcase success rates, as well as for token usage. To allow for long-term comparison, we suggest providing the estimated costs in US\$. Include an overview of executed command categories, frequently executed commands, and their error rates. We recommend the inclusion of qualitative analysis but cautiously suggest introducing a qualitative methodology for these within papers. We would prefer more advanced metrics but acknowledge that these typically involve time-consuming manual qualitative analysis. If feasible, executed commands and their errors should be subject to a qualitative analysis. If the benchmark supports sub-tasks, these should be analyzed for progression rates and potential dead-ends.


\bibliographystyle{plainnat}
\bibliography{bibliography}

\end{document}